# Finger Based Technique (FBT): An Innovative System for Improved Usability for the Blind Users' Dynamic Interaction with Mobile Touch Screen Devices

Mohammed Fakrudeen, Sufian Yousef, and Mahdi H. Miraz

*Abstract* —**This paper presents Finger Based Technique (FBT) prototypes, a novel interaction system for blind users, which is especially designed and developed for non-visual touch screen devices and their applications. The FBT prototypes were developed with virtual keys to be identified based on finger holding positions. Two different models namely the single digit FBT and double digit FBT were propounded. FBT technique were applied using two different phone dialer applications: a single digit virtual key for the single digit FBT model and a double digit virtual key with audio feedback enabling touch as input gesture for the later one. An evaluation with 7 blind participants showed that single digit FBT was significantly faster and more accurate than double digit FBT. In addition to that, single digit FBT was found to be much faster than iPhone VoiceOver entry speeds in performing similar tasks. Furthermore, our research also suggests 11 accessible regions for quick access or navigation in flat touch screen based smart phones for blind users. These accessible regions will serve as a usability design framework and facilitate the developers to place the widget for the blind user for dynamic interaction with the touch screen devices. As far as is known to the authors, this is a novel suggestion.**

## I. INTRODUCTION

AS technology changes rapidly, the touch screen devices became dominant in the market. The keyboard is used as an input device to access the information on the desktop and laptop computers by the blind user using assistive technologies such as screen readers. However, such facilitations are not present in touch screen devices. Thus, the accessibility of information and technology becomes a major issue for the blind users. Leading organizations serving the blind people such as the American Foundation for the blind[1], the Royal National Institute of Blind people[2] and the National Federation of the blind[3] recommend iPhoneas the only touch screen device to be used by the blind people. They are also of the opinion that the complex setup process is involved to set accessibility features in the Android device.

The review of the related literature reveals the advancement made in touch screen accessibility for blind users. In spite of this, the basic task of text entry remains error prone and slow. An experiment conducted by Bonner *et al.*[4] report that the mean entry speed achieved by the blind users using VoiceOver on iPhone device is 0.66 WPM (Words Per Minute). Oliver *et al.* [5, 6] report that the mean text entry with a VoiceOver like input method was 2.1 WPM. Thus, an efficient text entry method is required for the blind users to perform eyes-free text input.

While VoiceOver technique involves exploration of desired virtual keys through hit and trial process, this paper propose a novel technique based on FBT to easily identify the virtual key on the flat touch screen surfaces.

Kane *et al.* [13] is of the opinion that the blind users do not need to understand the spatial representation of the interface layout, they simply need to perform the multi-touch gestures to access the information. However, it requires a lot of efforts from the blind user to understand the gesture to be performed indifferent applications. It is not only time consuming but also causes fatigue, which leads to reduce the performance of the blind users. So far, enough research has not been performed to eliminate or reduce the existing navigational complexity using current technologies such as VoiceOver.

The major contributions of this research include designing, developing and evaluating both the prototypes based on FBT technique. To implement the prototypes, the researchers of this project took advantages of the existing phone dialler application which is commonly being used by most users. The objectives of the present study are: 1) to identify the regions on the touch screen surface, which can be easily accessible by the blind users and 2) to analyse the entry speed of the blind users using the prototypes developed. The output of the research will enhance the understanding of placing widget in the accessible regions for built-in applications.

## II. BACKGROUND

In the Human-Computer Interaction (HCI) literatures, some non-visual touch screen text entry methods based on Braille have been proposed. BrailleType[6] is one of the

Manuscript received March 14, 2014; revised April 8, 2014.
Mohammed Fakrudeen is with the Anglia Ruskin University, Chelmsford, UK and also with Department of Computer Science and Software Engineering, College of Computer Science and Engineering, University of Ha'il, PO Box 2440, Ha'il, Saudi Arabia (e-mail: m.fakrudeen@uoh.edu.sa).
Sufian Yousef is with the Department of Engineering and the Built environment, Anglia Ruskin University, Chelmsford, UK (e-mail: sufian.yousef@anglia.ac.uk)
M. H. Miraz is with the Department of Computer Science and Software Engineering, College of Computer Science and Engineering, University of Ha'il, PO Box 2440, Ha'il, Saudi Arabia and also with Glyndŵr University, Wrexham, Wales, UK (e-mail: m.miraz@uoh.edu.sa and m.miraz@glyndwr.ac.uk).

Braille-based text entry method, where the dots in Braille cell is represented by splitting the screen into six virtual keys. The users type a character based on the raised dots in the character by selecting the specified region. Oliver *et. al.*[5] found that BrailleType has an entry rate of 1.45 WPM, which was significantly slower than VoiceOver technique.

Frey et al[7] proposed a text entry method like a Braille with six large dots(virtual keys) namely the BrailleTouch,. Using this method, for typing a character, the user has to input all the dots simultaneously. However, no evaluation was conducted to assess the performance of BrailleTouch. Similarly, to input a character, three simple gestures has to be performed in TypeIn Braille[8]. No evaluation was performed for TypeInBraille as well.

The HCI literatures also proposed text entry methods which are not based on Braille. ,To enter a character using No-Look Notes[4], two virtual keys must be selected. Their comparative evaluation of No-Look Notes with VoiceOver found that the entry speed were 0.66 WPM and 1.32 WPM respectively. Similarly, Sanchez and Aguayo[9] proposed virtual keys for text entry methods. However, there technique also remained unassessed as no evaluation was performed so far.

Oliver *et al.* [5]present NavTouch using which the user has to navigate through alphabets by performing sequence of gestures to enter a character. The entry rate was 1.72 WPM and an error rate was above 10% which is much slower than VoiceOver technique.

Tinwala and Mackenzie [10] and Yfantidis and Evreinov [11] have conducted text entry method through gestures with blind folded sighted users. Their evaluation does not apply to blind people since the input ability of sighted person varies significantly from that of a blind person.

As mentioned earlier, iPhone is found to be most common touch screen device used by blind user and recommended by Blind organization. In iPhone, VoiceOver technique is used to access the information by the blind user. When VoiceOver is running, a user can touch a virtual key to hear its audio feedback about the digit selected and double tap to activate that digit or a letter. Thus, VoiceOver enables eyes-free entry on the virtual QWERTY keyboard of the iPhone touch screen.

One of the important observations of this study is that it is imperative for developers to identify the accessible regions for the blind users. This will facilitate them to perform dynamic interactions as the sighted users do. The existing related technologies for blind navigation lack this phenomenon. Without understanding the accessible regions and their access speed for flat touch screen surfaces, it will be difficult for developers to place the widget for accessing and to build the effective system in the future. This is, in fact, one of the key motivations for conducting this research study.

### III. RATIONALE

Our approach is to reduce the exiting inconsistence between the action of pointing and the user's hand size. FBT technique sets reference points near to the position held by the hand. This method makes it easier for the blind users to identify the target. The FBT prototypes aim to enhance usability in terms of users' preferences, users' ability, and various screen sizes such as tablets.

In VoiceOver technique, the user has to explore to get desired key using audio feedback, and then activate the key. These two step processes increase the text entry speed. Some techniques use large virtual keys for quick exploration [9]but only few keys fit on the screen, as keys get larger. In contrast, FBT technique is aimed to perform faster as the users can interact using predefined identifiable regions.

Finally, the chording nature and algorithm were used to make the text entry easier. Blind users observe difficulties if complex gesture is used as input. However, FBT technique adopts simple touch gesture for text entry. This approach also allows the users to correct text input errors, and to read what have been typed.

Current technologies such as iPhone and android touch screen smart phones implements a static layout for the interaction. In these techniques, two dimensional pages are collapsed to single dimensional to form a single horizontal list that contains a large set of items. Memorizing the larger sequence of interest items is additional burden to the blind user.

By and large, the sighted users prefer dynamic interactions by identifying items using vision. On the contrary, the blind user has to flicker the static layout to identify the items. Thus, the relative position of item such as on the top, on the bottom is lost. . In addition to that, developers have to incorporate separate navigation for the blind users. Furthermore, blind users' usability dimensions and requirements differ from those of the sighted users.

Thus, this study of designing, developing and evaluating the prototypes developed for the blind people using FBT for digital text entry is authentic and timely

### IV. FBT TEXT ENTRY METHOD

The efficient use of touch screen smart phones by blind users depends on the following factors:

1) The orientation/positions of the buttons
2) Voice feedback upon tapping/touching the buttons
3) Number of buttons included in the design and
4) The availability of required operations for correcting any mistakes committed.

The innovative prototype of FBT on touch screen based smart phones consists of two models: Single Digit FBT(as shown in Fig. 1) and Double Digit FBT(as shown in Fig. 2). The Single Digit FBT consists of virtual keys with single digit. On the other hand, each virtual key corresponds to two different digits in the Double Digit FBT. Both models take touch/tap as input and provides voice feedback upon 'virtual key press' action or if any mistake is committed by the blind users.

In both cases, as shown in Fig. 1 and Fig. 2, the user holds the device in the left hand. The index finger, the middle finger, the ring finger and the little finger of left hand holds on one side of the device. The other side of the device is hold by the left hand thumb finger. The user taps the button using right hand finger(s).

In single FBT technique, as shown in Fig. 1, the blind user assumes that the regions near to the tips of left hand finger represent the associated virtual digit. For instance, if the user presses the region near to ring finger, virtual key representing the 3<sup>rd</sup> digit is invoked. Table 1 shows the list of virtual digital keys along with their respective finger holding positions.

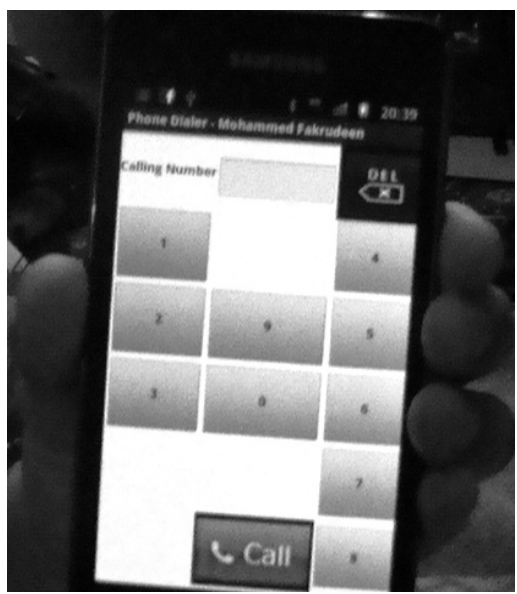

Fig. 1. Single Digit FBT

In double digit FBT, as shown in Fig. 2, the blind user assumes region vicinity to the holding tip of each finger represent the double digit. A single press will temporarily add the first digit to dialling number textbox. For adding the second number instead, two subsequent presses are required.

TABLE I
POSITION OF WIDGETS IN FBT

| Position of buttons with respect to holding left hand finger | Single Digit FBT | Double Digit FBT |
|---|---|---|
| Above index finger | Backspace | Backspace |
| Index finger | Four | One and Two |
| Middle finger | Five | Three and Four |
| Ring finger | Six | Five and Six |
| Little finger | Seven | Seven and Eight |
| Below Little finger | Eight | Nine and Ten |
| Centre of touch screen | Nine | |
| Thumb | Two | Enter |
| Above the thumb | One | |
| Below the thumb | Three | Contacts |
| Bottom Center of the touch screen | Call | Call |

For instance, to add "2" to the dialling number textbox, the blind users will press the region near to the tip of index finger to invoke "12" button. Upon the first press, the audio informs the users that "1" has been pressed. Upon the second press, the audio informs the users that "2" has been pressed. Now the users have to press the "Enter Key" to finally add "2" to the dialling number textbox. If the users want to add "1" instead, they have to press the "Enter Key" immediately after the pressing the "12" button for the first time.

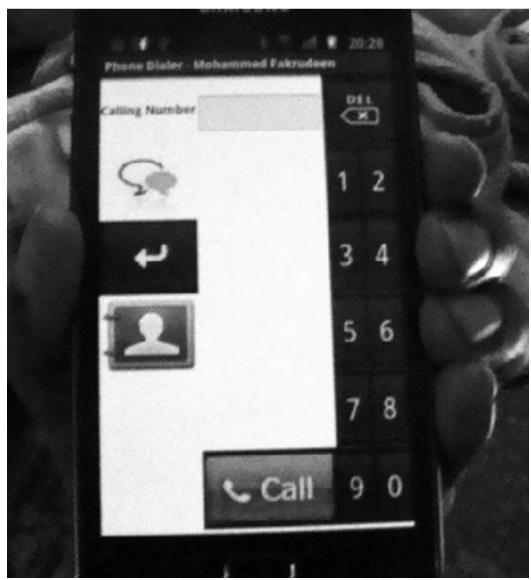

Fig. 2. Double Digit FBT

V. USER EVALUATION

The user study was carried out to evaluate the proposed technique for the blind users using the touch screen based smart phone.

*A. Participants*

With an average age of 35 years, we recruited seven blind participants. None of the participants had previous experience on using a touch screen mobile. However, they had enough experience of using mobiles with screen readers. Due to small cohort size, we adopted the "repeated trial" methodology and conducted 8 trials per participants.

*B. Equipments*

The FBT prototypes were developed using android 4.0 Ice Cream Sandwich. Both the prototypes can run on any mobile phone which supports android platform. Prototypes were tested on Samsung Galaxy S2. No additional hardware is required to run them. During the development phase, the prototypes were tested on emulator supplied by android SDK. After successful completion of the development phase, they were deployed on the Samsung Galaxy S2 mobile phone and tested with the blind users.

*C. Procedure*

The study includes two sessions: training and evaluation. During the training session, the participants were shown how to use the prototypes explaining the difference between the single digit FBT and double digit FBT.

The participants were then asked to perform small tasks such as dialling a number. After being able to call the first given number, the blind user were provided with another number to dial in order to assure the level of comfort.

Suring the evaluation session, each participant was given eight different contact numbers to dial. Each contact number consisted of ten digits. Since participants were blind, we were unable to present contact number visually. To overcome the cognitive load, simple number were provided so that the participants can remember them easily. To keep it simple and easy, one contact number was provided at a time i.e. just before performing each dialling task.

VI. ANALYSIS AND FINDINGS

The analysis conducted was based on: 1) the text to be entered by participant called Presented text (P) and 2) the text entered by participants called Transcribed text (T). This also involved the number of wrong digits entered and number of corrections performed by the participants. For both single and double digit FBT, the experiment was tested for the following factors:

1. Interface design
2. Preferences
3. Entry Speed
4. Error Rates

The first two factors were evaluated using the Likert Scale [1-Strongly Disagree, 5-Strongly Agree] based on the responses collected through questionnaires and data related to the other factors were recorded and collected during evaluation process.

*A. Words per Minute(WPM)*

WPM measures the time taken to produce target digit. As our aim of our research is not to build the messaging system but to adopt the process of text entry, we assume digit entry and text entry are to be similar process. Thus, the WPM, text entry metric is adopted for our technique. The WPM does not consider the gesture, the length of target number and the number of keystrokes made during digit entry. WPM is defined as[12]:

$$WPM = [(|T| - 1) / S] * 60 * (1/5)$$

Here, S is the time taken to enter the target number in seconds and |T| is the length of targeted number.

The examination of entry speed assesses the ability of blind users in accessing the location point according to finger positions.

WPM is checked for normality by using Shapiro Wilkson (W) test technique. The data are normalized for single digit FBT using WPM (W (6)=0.972, P<0.05) and double digit FBT (W(6)=0.896, P<0.05). Hence one way ANOVA was performed for normalized data to find the significance of each technique on the WPM.

Based on the result obtained it can be concluded that there is a statistically significant difference between single digit and double digit FBT as determined by one-way ANOVA ($F(1,10) = 6.600, p = .028$). It can be further concluded that the text entry speed (WPM) of Single digit (Mean=3.26, SD=0.784) was faster than that of double digit technique (Mean =1.92, SD=1.00) as shown in Fig. 3.

The achieved entry speed of the double digit FPD is slower as compared with single digit FPD owing to:

1. Participants have to press the button two times for digits such as 2,4,6,8,0 and
2. Participants have to press the enter key for confirmation (to add entry to the textbox).

After analysing the results, it can be concluded that single digit FBT technique is more promising than the double digit FBT with higher text entry rate.

*B. Performance over Duration*

The duration can be defined as the time takento complete dialing a phone number. The phone numbers provided consisted of 10 digits each. The entry speed was examined over the entire duration of the test.

The duration is checked for normality by using Shapiro Wilkson (W) test technique. The duration is normalized for single digit FBT using duration (W (6)=0.972, P<0.05) and double digit FBT using duration (W(6)=0.896, P<0.05). Therefore, one way ANOVA was performed to find the significance of each technique on the duration.

There exists statistically significant difference between single digit and double digit FBT as determined by one-way ANOVA ($F(1,10) = 4.499, p <0.05$). As a result, we can conclude that the time taken for text entry using Single (Mean=35.00, SD=9.67) was lesser than the time taken of double digit technique (Mean =68.50, SD=37.45).

*C. Error Count*

The number of digit(s) entered incorrectly was also analysed Out of seven participants, only two of them had to perform digit correction while using double digit FBT.

The error count is checked for normality by using Shapiro Wilkson (W) test technique. The number of error is not normalized for double digit FBT (W (6)=0.701, P=0.006). Hence Mann-Whitney U test was performed for not normalized data to find the significance of each technique on the error count.

There was no statistically significant difference between the error count of the techniques (U = 12, P =0.14). As a result, a firm conclusion cannot be made and hence it demands further research.

*D. Preference*

Although the entry speed of single digit FBT was faster than double digit FBT, 70% of the participants prefer double digit FPD as compared to Single Digit FBT (as shown in Fig. 4).

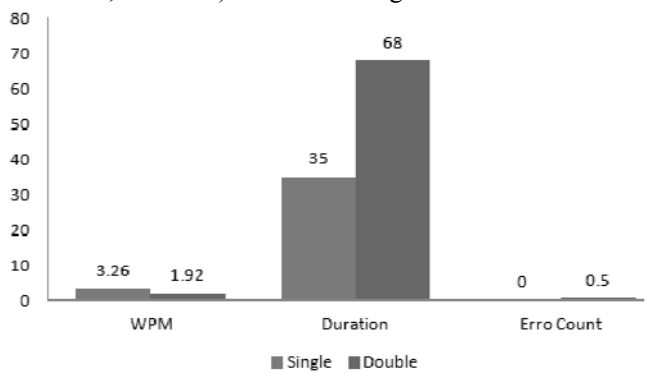

Fig. 3. Mean of dependent variables

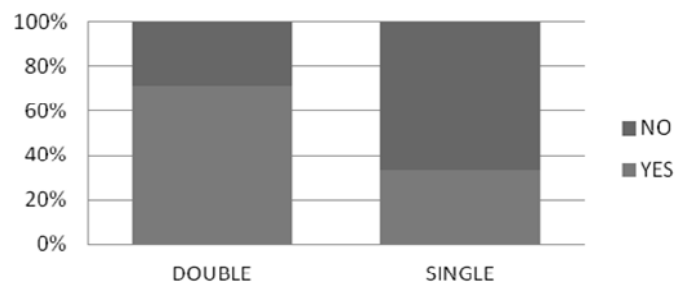

Fig. 4. Desires' of FBT

This was attributed to the following facts (as shown in Fig. 5):

1. *Location problem*: Only 5 buttons are placed in 'double digit' FBT for digits as compared to 10 buttons for 'single digit' FBT. Hence 60% of the participants found it to be difficult to locate the buttons in 'single digit' FPD.

2. *Time Consuming*: 40% of blind users are of the opinion that the 'double digit' FBT is time consuming as compared to 'single digit' FBT because:

    a. They had to press the same button twice for certain digit such as 2,4,6,8 and 0.

    b. They also had to press the enter key for confirmation of adding each digit.

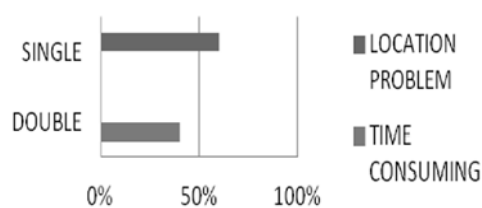

Fig 5. Problem related to FBT

## VII. DISCUSSION

Initially, during training session, performance of the participant was not as good as expected. Participants had difficulty remembering the calling number. The main challenge faced by the participants during the study was the lack of experience with touch screens. However, the training session improved their performance at a satisfactory level. The inclusion of the training session is hence justified.

The mean entry speed of single digit FBT is 3.26, which is higher than the mean entry speed 2.1 WPM, in iPhone VoiceOver by Oliver *et al*. In double digit FBT, the mean entry speed is 1.93, which is lower than Oliver's VoiceOver entry speed. The VoiceOver entry speeds in related work [4, 5, 6] are even lower than our findings. As a result, the single digit FBT outperforms VoiceOver, the *de facto* eyes-free text entry method for touch screen devices.

TABLE II
POSSIBLE ACCESSIBLE POINTS IN TOUCH SCREEN

| Location # | Position |
|---|---|
| 1 | Above index finger |
| 2 | Index finger |
| 3 | Middle finger |
| 4 | Ring finger |
| 5 | Little finger |
| 6 | Below the little finger |
| 7 | Above the thumb |
| 8 | Thumb |
| 9 | Below the thumb |
| 10 | Between thumb and middle finger |
| 11 | Bottom centre |

Our research provides insight into the performance of our technique by blind user using touch screen devices. The study shows that single digit FBT is much faster than double digit FBT. In addition, blind users helped to identify 11 reference regions for quicker accessibility and navigation for flat touch screen device. Identification of these accessible regions will enable the developers to position the widgets at the appropriate places for better user interaction.

Our technique relies on the marks at the edge of the touch screen where blind user holds the device. In reality, the overlays or mark can be seen in ATM machine to facilitate the sighted user to press the button easily at overlay position in flat touch screen machine. We tested both of our applications using different devices such as tablet. Interestingly, all the buttons lies on the edge at finger holding position. Thus, our application hold promises for cross-platform compatibility. However, it needs more analysis and testing.

Our findings are specific to smart phones and other mobile devices. These findings give reliability in our techniques for accessing the flat touch screen without any additional assistive device.

## VIII. CONCLUSION

Our work suggests an innovative interaction technique for flat touch screen to be used by the blind users. The proposed technique facilitates the developers to place the widget for the blind user for dynamic interaction with the touch screen devices. The evaluation shows that our technique is much faster than iPhone VoiceOver entry speeds in performing similar tasks.

The paper also focused on identifying the accessible regions for dynamic interaction. By properly adopting these techniques, non-functional requirement such as system usability can be achieved.

REFERENCES

[1] D. Burton, "Cell Phone Access: The Current State of Cell Phone Accessibility," 2011. [Online]. Available: http://www.afb.org/afbpress/pub.asp?DocID=aw120602.

[2] RNIB, "A guide to talking mobile phones and mobile phone software," [Online]. Available: http://www.rnib.org.uk/livingwithsightloss/Documents/Mobile_phone_software_factsheet_PDF.pdf. [Accessed 02 03 2014].

[3] NFB, "How many children in America are not taught to read?," 2013. [Online]. Available: https://nfb.org/braille-initiative. [Accessed 05 03 2014].

[4] M. N. Bonner, J. T. Brudvik, G. D. Abowd and W. K. Edwards, "No-look notes: accessible eyes-free multi-touch text entry," in *Proceeding Pervasive'10 Proceedings of the 8th international conference on Pervasive Computing*, 2010.

[5] J. Oliveira, T. Guerreiro, H. Nicolau, J. Jorge and D. Gonçalves, "Blind people and mobile touch-based text-entry: acknowledging the need for different flavors," in *ASSETS '11 The proceedings of the 13th international ACM SIGACCESS conference on Computers and accessibility*, Dundee, Scotland, 2011.

[6] J. Oliveira, T. Guerreiro, H. Nicolau, J. Jorge and D. Gonçalves, "BrailleType: unleashing braille over touch screen mobile phones," in *INTERACT'11 Proceedings of the 13th IFIP TC 13 international conference on Human-computer interaction - Volume Part I*, 2011.

[7] B. Frey, C. Southern and M. Romero, "Brailletouch: mobile texting for the visually impaired," in *UAHCI'11 Proceedings of the 6th international conference on Universal access in human-computer interaction: context diversity - Volume Part III*, 2011.

[8] S. Mascetti, C. Bernareggi and M. Belotti, "TypeInBraille: a braille-

based typing application for touchscreen devices," in *ASSETS '11 The proceedings of the 13th international ACM SIGACCESS conference on Computers and accessibility*, Dundee, Scotland, 2011.

[9] J. Sánchez and F. Aguayo, "Mobile messenger for the blind," in *ERCIM'06 Proceedings of the 9th conference on User interfaces for all*, Bonn, Germany, 2006.

[10] H. Tinwala and S. S. Mackenzie, "Eyes-free text entry on a touchscreen phone," in *IEEE Toronto International Conference Science and Technology for Humanity (TIC-STH)*, Toronto, 2009.

[11] G. Yfantidis and G. Evreinov, "Adaptive blind interaction technique for touchscreens," *Universal Access in the Information Society,* vol. 4, no. 4, pp. 328 - 337, 2006.

[12] R. W. Soukoreff and I. S. MacKenzie, "Metrics for text entry research: an evaluation of MSD and KSPC, and a new unified error metric," in *CHI '03 Proceedings of the SIGCHI Conference on Human Factors in Computing Systems*, Fort Lauderdale, Florida, 2003.

[13] S. K. Kane, J. P. Bigham and J. O. Wobbrock, "Slide rule: making mobile touch screens accessible to blind people using multi-touch interaction techniques," Halifax, Nova Scotia, Canada, 2008.